\documentclass{ws-ijmpd}
\begin{document}

\markboth{Meng, Gao \& Han} {SNe Ia as a cosmological probe}

%
\catchline{}{}{}{}{}
%

\title{SNe Ia as a cosmological probe}

\author{Xiangcun Meng\footnote{This article is to be published
also in the book ¡°One Hundred Years of General Relativity: From
Genesis and Empirical Foundations to Gravitational Waves,
Cosmology and Quantum Gravity,¡± edited by Wei-Tou Ni (World
Scientific, Singapore, 2015).}}

\address{BPS Group, Yunnan Observatories, Chinese Academy of Sciences, \\
Kunming, 650216, China\\
xiangcunmeng@ynao.ac.cn}

\author{Yan Gao}

\address{BPS Group, Yunnan Observatories, Chinese Academy of Sciences, \\
Kunming, 650216, China\\
ygbcyy@ynao.ac.cn}

\author{Zhanwen Han}

\address{BPS Group, Yunnan Observatories, Chinese Academy of Sciences, \\
Kunming, 650216, China\\
zhanwenhan@ynao.ac.cn}

 \maketitle

\begin{history}
\received{19 July 2015} \revised{12 August 2015}
\end{history}

\begin{abstract}
Type Ia supernovae luminosities can be corrected to render them
useful as standard candles able to probe the expansion history of
the universe. This technique was successful applied to discover
the present acceleration of the universe. As the number of SNe Ia
observed at high redshift increases and analysis techniques are
perfected, people aim to use this technique to probe the equation
of state of the dark energy. Nevertheless, the nature of SNe Ia
progenitors remains controversial and concerns persist about
possible evolution effects that may be larger and harder to
characterize than the more obvious statistical uncertainties.
\end{abstract}


\ccode{PACS numbers:}


\section{Introbuction}\label{sect:1}
``Guest stars'', the name given to supernovae by ancient Chinese
astronomers, have been recorded for several thousands of years,
but the first modern study of supernovae began on the 31 August
1885. On that day, Hartwig discovered a ``nova'' in M31, which
disappeared 18 months later. In 1919, Lundmark\cite{LUNDMARK1920}
noticed that the distance to M31 is about 0.7 million ly (later
studies found it to be actually 2.5 million ly), which implies
that the Hartwig ``nova'' is brighter than other novae by three
orders of magnitude. The term ``supernova'' was coined shortly
after. Later, it was realized that supernovae can be a powerful
tool for measuring the distance to extragalactic sources. However,
for want of detailed enough observed spectra, supernovae were not
classified until 1940\cite{MINKOWSKI40}. With the advent of
consistently accurate observed spectra, it was finally realized
that there are two physically distinct classes of supernovae:
thermonuclear supernovae, i.e. SNe Ia, and core collapse
supernovae, including SN IIP, IIL, IIn, IIb, Ib and
Ic\cite{FILIPPENKO97}. Type I supernovae are distinguished by the
absence of hydrogen lines in their optical spectra. SNe Ia are
defined by a deep absorption trough around 6150 ${\rm \AA}$, which
is now known to be due to silicon,
additionally\cite{FILIPPENKO97}.

Due to the homogeneity of their properties, SNe Ia are often used
to measure distance. Here, we will review the methods commonly
used to standardize SNe Ia for this purpose in Section.
\ref{sect:2}, and the progenitor problem in Section. \ref{sect:3}.
In \ref{sect:4}, we discuss the effect of different populations of
SNe Ia on their brightness. In Section. \ref{sect:5}, we will
present the application of SNe Ia in cosmology, i.e. the discovery
of accelerating cosmic expansion, and the efforts to probe the
equation of the state of dark energy. Finally, we summarize the
future promise for SNe Ia as the probe to precision cosmology in
Section. \ref{sect:6}.

\section{SNe Ia as a standardizable distance candle}\label{sect:2}
\begin{figure*}
\begin{center}
\includegraphics[totalheight=2.5in,width=3.in,angle=0]{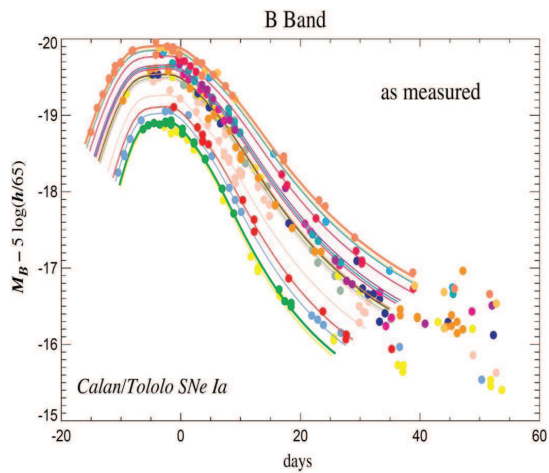}
 \caption{The B band lightcurve for different SNe Ia observed (From Kim 2004$^{\rm 9}$).}
 \label{curve}
  \end{center}
\end{figure*}

Generally, for two nearby objects ($z\ll1$) with astronomy
apparent magnitudes $m_{\rm 1}$ and $m_{\rm 2}$, so that a
Euclidean spatial geometry can be assumed, the ratio of their
apparent brightness, $E_{\rm 1}/E_{\rm 2}$, is related by the
following equation,
 \begin{equation}
\frac{E_{\rm 1}}{E_{\rm 2}}=100^{\frac{m_{\rm 2}-m_{\rm 1}}{5}},
  \end{equation}
which indicates that if the apparent magnitudes of one object is
lower than that of another one by 5 magnitudes, the object is
brighter than another by 100 times. For the same object, its
apparent magnitudes at 10 parsecs is defined as absolute
magnitude, which may reflect its intrinsic luminosity. Assuming
the absence of extinction by dust, its apparent magnitude, $m$,
and absolute magnitude, $M$, follow the equation
 \begin{equation}
\frac{E_{\rm 10}}{E}=100^{\frac{m-M}{5}}=\frac{d^{\rm 2}}{10^{\rm
2}}, \label{disinsv}
  \end{equation}
where $E_{\rm 10}$ is the apparent brightness at 10 parsecs, $E$
the observed brightness, and $d$ the distance to the object in
units of parsecs. Equation (\ref{disinsv}) basically expresses the
inverse-square law between apparent brightness and distance.
Therefore, apparent magnitude, absolute magnitude and distance are
related by the following equation,
 \begin{equation}
M=m+5-5\log d, \label{distance}
  \end{equation}
and $m-M=5\log d-5$ is called distance modulus. As we shall
discuss later in Section. \ref{sect:5}, at larger redshifts, for
which $z\ll1$ does not hold, the relation between apparent and
absolute magnitude is no longer so simple and depends on the
details of the cosmological model assumed. If there is a method to
independently obtain the absolute magnitude of an object, the
distance may be obtained by measuring its apparent magnitude, like
equation (\ref{distance}), and then this object may be taken as a
standard candle to measure distance. SNe Ia are excellent
candidates for standard candles after corrected by some methods.

\begin{figure*}
\begin{center}
\includegraphics[totalheight=3.in,width=3.in,angle=0]{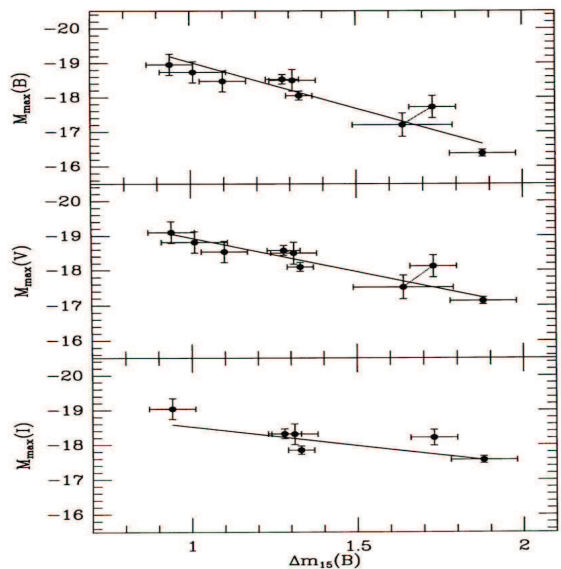}
 \caption{The correlation between the absolute magnitude of SNe Ia in B, V and I bands
 and $\Delta m_{\rm 15}(B)$ which measures the B band light-curve drops during the first 15 days after
 the maximum light (From Phillips 1993$^{\rm 10}$).}
 \label{phillips1}
  \end{center}
\end{figure*}

\begin{figure*}
\begin{center}
\includegraphics[totalheight=3.in,width=3.in,angle=0]{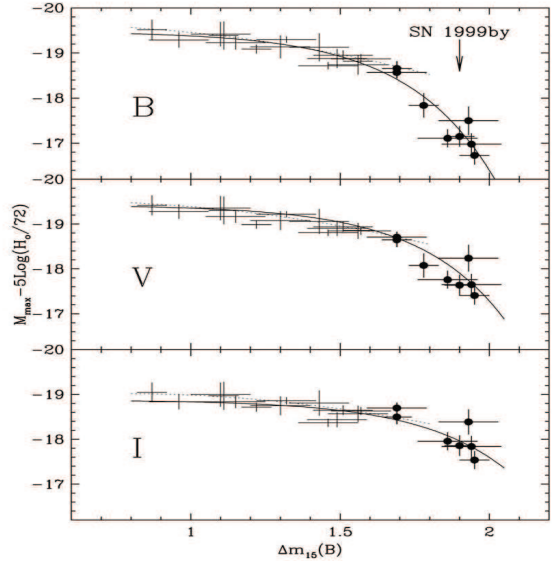}
 \caption{Modified Phillips' relation with so-called peculiar SNe added (filled points).
 The solid line is an exponential fit for all the data (From Garnavich et al. 2004$^{\rm 12}$).}
 \label{phillips2}
  \end{center}
\end{figure*}

Among all the sub-classes of different
supernovae\cite{FILIPPENKO97}, SNe Ia are the most homogeneous,
with many practically identical properties, e.g. light curve
shape, maximum luminosity, spectrum, and as a result, SNe Ia were
taken to be perfect standard candles\cite{BRANCH92}. However, 1991
was a fateful year for SNe Ia studies, two peculiar SNe Ia were
found, SN 1991bg and 1991T. SN 1991bg was fainter than normal SNe
Ia by about 2 magnitudes in the V
band\cite{FILIPPENKO92a,LEIBUNDGUT93}, while SN 1991T was brighter
than normal SNe Ia by 0.4
magnitudes\cite{FILIPPENKO92b,PHILLIPS92}. The discovery of these
two peculiar SNe Ia implies a distance error of about 30\% if SNe
Ia are assumed to be perfect standard candles. As shown by the
collection of different supernova lightcurves in Fig. \ref{curve},
different SNe Ia have different peak brightnesses. It became a
matter of critical importance to find a way to reduce the distance
error originating from this heterogeneity of SNe Ia\cite{KIM15}.
In 1993, Phillips\cite{PHILLIPS93} discovered that the absolute
magnitude at maximum light of SNe Ia and the speed at which the
luminosity fades in the B-band (blue light) over the 15 days after
the maximum light are related, as shown in Fig. \ref{phillips1}.
This relation implies that the brightness of SNe Ia is mainly
dominated by one parameter, and it is widely agreed that this
parameter is the amount of $^{\rm 56}$Ni produced during the
supernova explosion that determines its maximum luminosity.
Actually, one may arrive at another conclusion from Fig.
\ref{phillips1}, that the intrinsic magnitude dispersion of SNe Ia
in the I band is smaller than those in the B and V bands, i.e. the
infrared measurement of SNe Ia may yield a more precise distance.
However, several years later, after increasingly dim SNe Ia were
included, the linear relation was found to be a quadratic or an
exponential relation as shown in Fig.
\ref{phillips2}\cite{PHILLIPS99,GARNAVICH04}. Although this
relation is widely accredited to Phillips, it was originally
discovered by Rust\cite{RUST74} and
Pokovskii\cite{PSKOVSKII77,PSKOVSKII84}, who noticed the
correlation between the maximum light of SNe Ia and their light
decline rate.

\begin{figure*}
\begin{center}
\includegraphics[totalheight=3.in,width=3.in,angle=0]{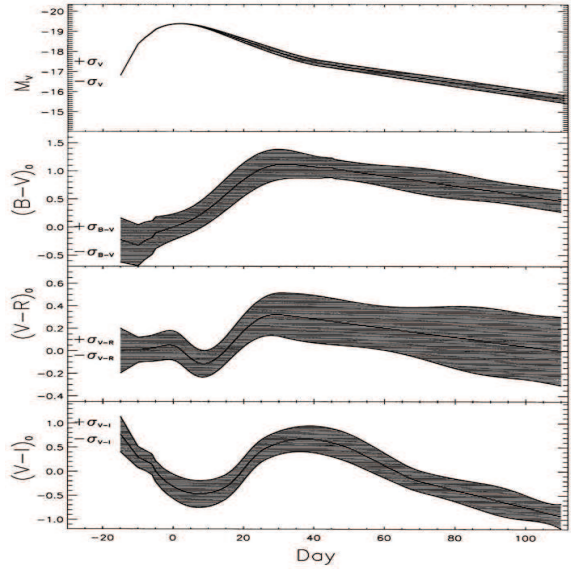}
 \caption{Dispersions of noise-free light and color curves after correction by MLCS
 method, where the ``gray snake'' present 1 $\sigma$ confident
 region (From Riess et al. 1996$^{\rm 16}$).}
 \label{mlcs}
  \end{center}
\end{figure*}

\begin{figure*}
\begin{center}
\includegraphics[totalheight=2.5in,width=3.in,angle=0]{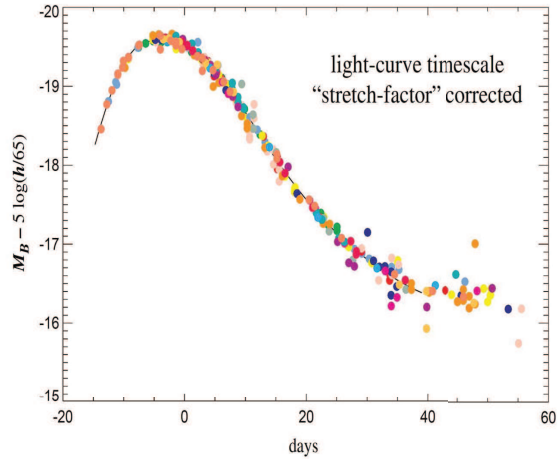}
 \caption{The lightcurve for different SNe Ia after correction by stretch factor method,
 where the data are the same to that in Fig. \ref{curve} (From Kim 2004$^{\rm 9}$).}
 \label{stretch}
  \end{center}
\end{figure*}

Following the discovery of Phillips\cite{PHILLIPS93}, several
groups developed further improved methods to reduce the distance
error, and SNe Ia were not a standard candle any more, but a
distance indicator. Most notably, Riess et
al.\cite{RIESS96,RIESS98} developed the mult-color light-curve
shapes(MLCS) method, which is known as the most mathematically
sound method for inferring a distance. The method account for the
non-uniform error on the distance from the effect of different
bands and different supernova colors. In this method, a ``training
set'' of several SN Ia light curves was constructed, and a
complete model of the light curves is dependent on three
parameters, i.e. a distance modulus, a brightness offset which is
called ``luminosity correction'', and an extinction value. To
calibrate the brightness of the training set, the distance modulus
and the extinction value are obtained by independent methods.
Actually, the brightness offset reflects that a SN Ia with a
broader light curve has a higher peak luminosity, i.e. the
Phillips relation. The reason that a correction was made for color
is that redder SNe Ia are generally less luminous, both in
intrinsic terms, and for dust reddening
considerations\cite{RIESS96,CONLEY07}. Compared to the Phillips
relation, although the MLCS method does not significantly increase
the precision of distance measurements, the method may be the most
complete and rigorous mathematically. Fig. \ref{mlcs} shows the
typical dispersions of light and color curves after correction by
the MLCS method, and from the figure, we can see that the SNe Ia
can be very good distance indicators, because the dispersion at
peak brightness is very small. Almost at the same time, Perlmutter
et al.\cite{PERLMUTTER97,PERLMUTTER99} developed another tool,
named the stretch factor method. This method has a similar
distance precision compared with the MLCS method, as shown in Fig.
\ref{stretch}, where the data are the same to those in Fig.
\ref{curve}. Again, SNe Ia are proven to be potent distance
indicators. It is worth noting that the MLCS method and the
stretch factor method essentially take advantage of the same
underlying phenomena as that underlying the Phillips relation,
i.e. the slower the lighcurve evolves, the brighter the SN Ia. To
obtain the parameters crucial to the implementation of these
methods, various algorithms have been developed to fit the light
curves of SNe Ia data, such as BATM, MLCS2k2, SALT2, and
SiFTO\cite{CONTARDO00,TONRY03,JHA07,GUY07,CONLEY08,DADO15}.

Other research groups have also tried to find other methods to
obtain more accurate SNe Ia distance measurements. For
example, in 1995, Nugent et al.\cite{NUGENT95} noticed a linear
relation between the maximum luminosity of a SN Ia and the ratio of
the fractional depth of the bluer to that of the redder absorption trough
of its silicon absorption line, or the flux ratio over
the Ca II, H and K absorption features in a spectrum near maximum
light (see also \cite{BAILEY11,SILVERMAN12}). SNe Ia were
therefore calibrated to be practical standard candles,
and have been applied successfully for cosmological
purposes, ultimately leading to the discovery of the accelerating
expansion of the Universe\cite{RIESS98,PERLMUTTER99}.

The above methods generally decrease the distance error to 5\% -
10\%. However, for a well-observed sample, the typical dispersion
on a Hubble diagram may be as small as 0.12 mags, indicating a
typical distance accuracy of 5\% for a single SN Ia. In 2005, Wang
et al.\cite{WANGXF05} found that the peak brightness of SNe Ia and
their $B - V$ colors at 12 days after peak B-band luminosity were
strongly correlated. Calibrating SNe using this correlation, the
typical error of the SNe Ia on a Hubble diagram can be reduced to
less than 0.12 mags. For a typical sample of SNe Ia with little
reddening from host galaxies, the scatter may drop further to only
0.07 mag, or 3\%-4\% in distance \cite{WANGXF05,WANGXF06}. Here,
it must be emphasized that for most cases, the observational error
is not mainly due to imperfections in the observational apparatus,
but rather from a lack of uninterrupted telescope time, i.e. it is
difficult to obtain a complete light curve for each single object.
Recently, in addition to the light-curves and the colors of the
SNe Ia, some spectra properties were found to be helpful for
improving the accuracy of derived distances to SNe
Ia\cite{BAILEY11,SILVERMAN12,MAEDA11,KIM13}.

For precision cosmology, within most redshift intervals,
systematic errors from the above correlations dominate over the
statistical errors, the latter of which can be dealt with by
increasing the size of an observational
sample\cite{KESSLER09,HICKEN09,AMANULLA10}. The dominant
systematic uncertainties for SNe Ia are survey-dependent. As far
as what the uncertainties are and how they can be improved are
concerned, one may refer to relevant detailed
reviews\cite{HOWELL11,LEIBUNDGUT01,ASTIER12}.

\section{Progenitors of type Ia supernovae}\label{sect:3}
In 1960, Hoyle \& Fowler\cite{HOYL60} suggested that a degenerate
white dwarf (WD) may ignite a runaway thermonuclear fusion in its
center, and the released energy may disrupt the whole WD,
resulting in a type Ia supernova (SN Ia) explosion.  Owing to a
series of breakthroughs on the subject since then, it has been
indisputably proven that all SNe Ia arise from such explosions of
carbon-oxygen white dwarfs (CO WDs) in binary systems, of which
the most substantial evidence is derived from the observation of
SN 2011fe \cite{NUGENT11}. For a CO WD to undergo such an
explosion, it must reach or exceed a mass of about 1.4
$M_{\odot}$, henceforth referred to as the Chandrasekhar mass
limit. The limit was named after Subrahmanyan Chandrasekhar, who
was the first person to work out that there was a maximum mass for
a white dwarf to be supported by electron degeneracy pressure. He
did this by combining quantum theory with relativity, and it was
the first time to make a quantitative prediction about the maximum
mass of WDs. In 1983, he was awarded the Nobel Prize in Physics
mainly due to the discovery of the mass limit of WDs.

However, the mass of a CO WD at birth cannot be more than 1.1
$M_{\odot}$\cite{UME99,MENG08}, according to canonical stellar
evolutionary theory, and therefore the CO WD must accrete
additional material somehow (i.e. from its companion in a binary
system), and gradually increase its mass to the aforementioned
maximum stable mass before it can achieve this state and undergo a
SN explosion. The details of this accretion process will be
discussed later. During a SN Ia explosion, nearly half of the mass
involved is depleted and synthesized into radioactive $^{\rm
56}$Ni, the radioactive energy of which is then injected into the
supernova ejecta, heating it to the point where it becomes very
luminous, ultimately resulting in the emissions which we
observe\cite{BRANVH04}. The amount of $^{\rm 56}$Ni resulting from
a SN explosion is the dominant determining factor behind the
maximum luminosity of the SN Ia\cite{ARNETT82}. However, despite
all that we know, there still remain quite a few unresolved
problems where SNe Ia are concerned. To name but a few, what are
the progenitors of SNe Ia, or how does a CO WD increase its mass
to the Chandrasekhar limit? How exactly does the WD explode? Do
the basic properties of SNe Ia remain constant with redshift?
Among all these basic problems, the most urgent is the one
regarding the progenitors of SNe Ia, which can potentially affect
the precision of distance measuring via SNe Ia, and hinder the
development of precision cosmology. More specifically, the
progenitor problem is important, for cosmology as well as for
other areas of astrophysics, in the following ways. 1) When
measuring the cosmological parameters with SNe Ia, one needs to
know of any possible evolution of the luminosity and birth rate of
SNe Ia as a function of redshift, which is mainly determined by
the progenitor model. 2) The progenitor model provides basic input
parameters for explosion models of SNe Ia. 3) Galactic chemical
evolution models require some basic parameter input, such as the
stellar feedback from SNe Ia, both chemical and radiative. These
basic input parameters are closely related to the progenitor
models of SNe Ia. 4) The identification of the progenitor systems
of SNe Ia may constrain binary evolution
theory\cite{LIVIO99,LEI00,WANGB12,MAOZ14}.

\begin{figure*}
\begin{center}
\includegraphics[totalheight=3.in,width=5.in,angle=0]{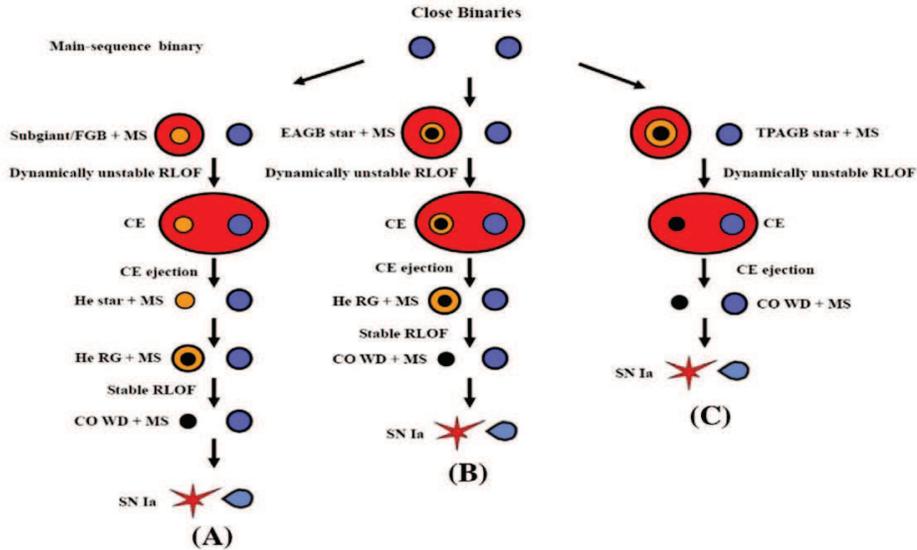}
 \caption{Schematic for the sub-channels to producing WD + MS systems, which may lead to SNe
 Ia. For every sub-channel, the binary system experiences a
 common envelope phase
 (From Wang \& Han 2012$^{\rm 48}$).}
 \label{ms}
  \end{center}
\end{figure*}

Over the course of the last 3 decades, several progenitor models
were discussed by many research groups. These models may be
categorized based on the nature of the proposed companion of the
CO WD, and we list the most notable ones below. 1) The single
degenerate (SD) scenario\cite{WI73,NTY84}: In this scenario, the
companion may be a main sequence star, a subgiant, a red giant or
a helium star\cite{LIXD97,HAC99a,HAC99b,LAN00,HAN04,WANGB09}. The
accreted hydrogen-rich or helium-rich material undergoes stable
burning on the surface of a CO WD, yields carbon and oxygen, which
is then deposited onto the WD. When the WD mass reaches some value
approaching the Chandrasekhar mass limit, a SN Ia is produced. The
companion survives after the supernova explosion.  This scenario
is widely accepted and is the leading scenario. 2) The double
degenerate (DD) scenario: In this scenario, a binary system
comprising two CO WDs loses its orbital angular momentum by means
of gravitational radiation, and merges as a consequence. If the
total mass of the merger exceeds the  Chandrasekhar mass limit, a
SN Ia may be expected\cite{IT84,WEB84,PAKMOR10,VANKERK10}. In this
scenario, the merger is disrupted completely and the companion
does not survive. Note, however, it is possible that the merger
product may experience a core collapse rather than a thermonuclear
explosion\cite{SAIO85,SAIO98}. The SD and DD scenarios are the two
main competing scenarios in the SN Ia community. Generally, to
form a SD system leading to a SN Ia, a primordial binary system
needs to experience one common envelope phase, while to form a DD
system, the situation is more complicated and one extra common
envelope phase can be necessary. For example, in Fig. \ref{ms},
three sub-channels lead to a WD + main sequence system ending in a
SN Ia explosion, and for every sub-channel, the system experiences
a single common envelope phase only (see details in Wang \&
Han\cite{WANGB12}).

In addition to the models mentioned above, the double-detonation
model (sub-Chandrasekhar mass model) is also frequently discussed.
In this model, the companion of the CO WD is a helium WD or a
helium star, which fills its Roche Lobe and initiates a stable
mass transfer. If the mass-transfer rate is not high enough, the
helium material may not burn stably, and as a consequence
gradually accumulates on the CO WD. When the helium layer is
massive enough, a detonation, which is a supersonic blast wave, is
ignited at the bottom of the helium layer, in which the critical
mass to ignite a detonation is generally dependent on the WD mass
and the mass transfer rate\cite{IT89}. At the moment of the
ignition, the WD is generally less massive than the Chandrasekhar
mass limit. The resulting detonation wave compresses the inner CO
material, leading to a second detonation near the core of the CO
WD\cite{WOOSLEY94,LIVIN95,SHENK13}. After the supernova explosion,
a low mass helium star or a low mass helium WD may survive to form
a hyper-velocity star, whose velocity is high enough to exceed the
escape velocity of its parent galaxy\cite{GEIER15}. Remarkable
progress has been made on this model recently, though many
problems still exist\cite{SHENK09,SIM12}. If this model does
indeed produce SNe Ia, it would contribute quite significantly to
SNe Ia rates\cite{RUITER11}.

Besides the SD and DD scenarios and the double detonation model,
other similar models have been proposed to explain the diversity
generally observed among SNe Ia, such as the the
super-Chandrasekhar mass model, the single star model, the
spin-up/spin-down model, the delayed dynamical instability model,
the core-degenerate model, the model of a collision between two
WDs, and the model of WDs near black holes, etc. (See Wang \& Han
\cite{WANGB12} for a review). At present, it is premature to
exclude any of these models conclusively, and no single model can
explain all the properties of SNe Ia alone. It is possible that
many of these models, or at least both the SD and DD scenarios,
contribute to SNe Ia\cite{WANGB12,MAOZ14,RUIZLAPUENTE13,WANGXF13}.

\section{Effect of SN Ia populations on their brightness}\label{sect:4}
When applying SNe Ia as cosmological standard candles, it is usually
assumed that their properties and the above calibrated relations
are invariant with redshift, i.e. the demographics of SNe Ia does
not evolve with redshift, or the difference between SNe Ia arising
from different stellar populations is negligible. This assumption
is crucial for precision cosmology, since it may lead to systematic
errors in the measurements of cosmological parameters. One way to
test this assumption is to search for correlations between the
properties of SNe Ia and those of their environments, since the environments
of SNe Ia can represent the host population which gives rise to
SNe Ia to a great extent. By the merit of these studies, it is now
known that SN light curves and peak luminosities are correlated
with host galaxy star formation rate, host population age, galaxy
mass and
metallicity\cite{SULLIVAN06,HOWELL09,GALLAGHER08,SULLIVAN10}.

Even before SNe Ia were applied to measure cosmological
parameters, it was generally agreed upon that the brightest SNe Ia
always occur in late-type galaxies\cite{HAMUY96}. It should be
noted, however, that both late- and early-type galaxies can host
dimmer SNe Ia, and sub-luminous SNe Ia tend to be discovered in
galaxies with a significant old population\cite{HOWELL01}, which
leads to a lower mean peak luminosity for SNe Ia in early-type
galaxies than in their late-type
counterparts\cite{HAMUY96,BRANDT10}. In addition, the mean peak
brightness of SNe Ia in a galaxy are more homogeneous in the outer
than in the inner regions\cite{WANGLF97,RIESS99}. Also, both the
birth rate and luminosity of SNe Ia trace the star formation rate
of host galaxies\cite{SULLIVAN10,SULLIVAN06}. All these
observations seem to imply that age is one of the factors which
affect the luminosity of SNe Ia, and that dimmer SNe Ia can arise
from progenitors spanning a wide range of
ages\cite{HOWELL09,GALLAGHER08,NEILL09,GUPTA11,RIGAULT15}.
Considering that typical galactic star formation rates increase by
a factor of $\sim10$ up to $z=1.5$, one may expect that the
luminosities of SNe Ia will generally increase with redshift, and
it has been shown that the mean ``stretch factor'' increases as
well\cite{HOWELL07,SULLIVAN09}. According to these observations,
we may conclude that the average value of the maximum luminosities
of SNe Ia, as well as the range over which they span, both
decrease with their delay time. It can also be concluded that,
should the maximum luminosity be determined by only one parameter,
as postulated by the Phillips relation\cite{PHILLIPS93}, it can be
expected that the possible range of the parameter must decrease,
and that its mean value must either increase or decrease with the
age of SNe Ia. At present, although it has been widely accepted
that the peak luminosity of SNe Ia is dictated by the quantity of
$^{\rm 56}$Ni produced during their explosion, it is still
controversial which parameter physically determines the $^{\rm
56}$Ni production\cite{PODSIADLOWSKI08}. The dimmer brightness for
an old SN Ia could be either due to a high ignition density for a
long cooling time before accretion\cite{KRUEGER10,CHENXF14}, or
due to a lower carbon abundance requirement for a massive WD with
a less massive secondary under the frame of the SD
model\cite{MENGXC10,UMEDA99b,MENGX11c}. In addition, the total
mass of the DD systems was suggested to the origin of the
brightness variation of SNe Ia\cite{HOWELL11,MAOZ12}, but it seems
not to fit the age constrains on the brightness of SNe Ia from
observations\cite{MENGXC12}.

Metallicity is another factor affecting the luminosity of SNe
Ia\cite{HOWELL09,GALLAGHER08,SULLIVAN10,CHILDRESS13,HAYDEN13}.
Theoretically, high metallicity progenitors would produce
sub-luminous SNe Ia: a high $^{\rm 22}$Ne abundance in high
metallicity white dwarfs generates more neutrons to feed the
explosion nucleosynthesis, yielding more stable $^{\rm 58}$Ni, a
process which consumes the radioactive $^{\rm 56}$Ni powering the
lightcurve of SNe Ia\cite{TIMMES03}. The spectra of high-z SNe Ia,
whose metellicity is on average lower than that of local SNe Ia,
contain less intermediate mass elements, which is consistent with
the idea that these SNe Ia produce more $^{\rm 56}$Ni to power
their luminosity\cite{SULLIVAN09}. Metallicity may lead to a
$\sim25$\% variation on the peak luminosity of SNe Ia, but
observation has shown that metallicity might only have a $\sim10$
\% effect on this luminosity\cite{HOWELL09}. Such a difference
between theory and observation could be the result of a
metellicity-age-carbon abundance degeneracy, or an unknown
metallicity effect. A high metallicity may lead to a relatively
shorter delay time\cite{MENGXC12,MENGX09,MENGX11b}, regardless of
what the progenitor model is, and hence a slightly brighter SNe
Ia. On the other hand, a star of given mass with a high
metallicity will produce a relatively less massive
WD\cite{UME99,MENG08} that has a higher carbon
abundance\cite{UMEDA99b}, which was suggested to produce a
brighter SNe Ia\cite{UMEDA99b,MENGX11c}. Therefore, the effect of
metallicity on the peak luminosity of SNe Ia could be more complex
than what one would imagine at first glance.

The mass of a host galaxy may also significantly affect the
properties of SNe Ia. For example, many observations have shown
that the birth rate of SNe Ia depends heavily on the mass of
their host galaxies\cite{MANNUCCI08,LIWD11b,GAOY13}. Compared with
the age and metallicity, the effect of the mass of host galaxies
could be more
significant\cite{SULLIVAN10,GUPTA11,CHILDRESS13,JOHANSSON13,RIGAULT13}.
More massive galaxies tend to give rise to dimmer SNe
Ia\cite{SULLIVAN10,RIGAULT13}, a trend which seems to be a
combined result of the effects of age and metallicity on the
brightness of SNe Ia, i.e. higher mass galaxies have the tendency to retain
more metals due to gravitational effects, and at the same
time, tend to have more old stars, supergiant
elliptical galaxies in particular.

\begin{figure*}
\begin{center}
\includegraphics[totalheight=2.5in,width=4.in,angle=0]{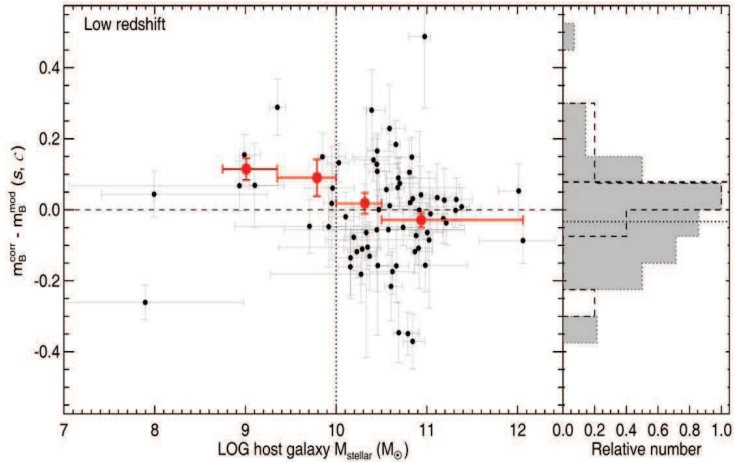}
 \caption{Residuals from the best-fitting cosmology for a
 low-redshift sample as a function of host galaxy mass after correction
 for light-curve shape and color. The red circles are the
mean residuals in bins of host galaxy mass. The right-hand
histograms show the distribution of the residuals for SNe Ia in
low (dashed histogram) and high (gray histogram) mass hosts, where
the mass boundary is $10^{\rm 10}$ $M_{\odot}$, and the horizontal
lines show the average residuals for the sub-samples in low and
high mass hosts (From Sullivan et al. 2010$^{\rm 77}$).}
 \label{sullivan}
  \end{center}
\end{figure*}

The above facts imply that the peak luminosities of SNe Ia evolve
with redshift. Despite this, if the absolute magnitude of SNe Ia
corrected by their lightcurve shape and color does not change with
redshift, it is not necessarily problematic to take SNe Ia as
standard candles for cosmology. However, people have to face an
embarrassing situation that the age, metallicity and host galaxy
mass have a systemic effect on the brightness of SNe Ia even after
their brightness was corrected by the lightcurve shape and
color\cite{SULLIVAN10,GUPTA11,CHILDRESS13,HAYDEN13,JOHANSSON13,RIGAULT13}.

For example, as shown in Fig. \ref{sullivan}, Sullivan et
al.\cite{SULLIVAN10} found that SNe Ia originating in massive host
galaxies and galaxies with low specific star-formation rates, of the
same light curve shape and color, are generally 0.08 mags brighter
than those found anywhere else. This result does not depend on any
assumed cosmological model or SN light-curve width. Such a trend may lead to a systematic
error on the measurement of the equation of state of dark energy,
if it is not corrected. Therefore, besides the lightcurve shape
and the color, at least a third parameter to correct the
brightness of SNe Ia, which is correlated to one or more
properties of the host galaxies, must be incorporated to avoid
systematic Hubble diagram residuals.

\begin{figure*}
\begin{center}
\includegraphics[totalheight=4.in,width=3.in,angle=0]{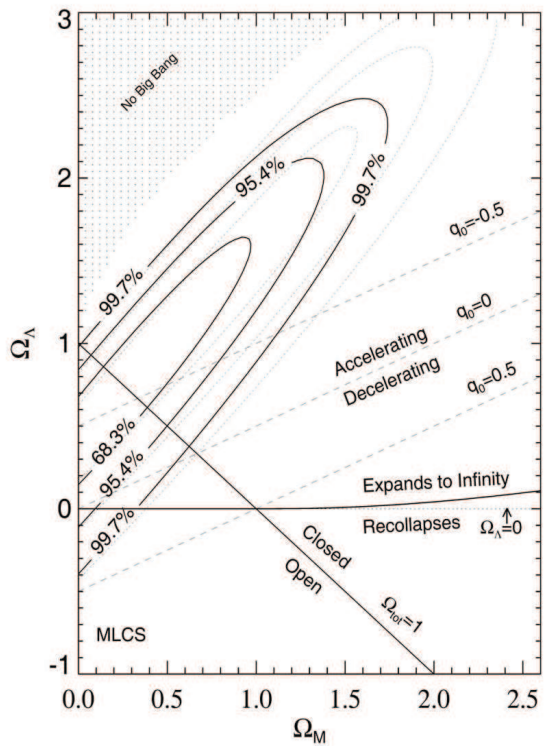}
 \caption{Joint confident level for ($\Omega_{\rm M}$, $\Omega_{\rm \Lambda}$) from
 SNe Ia corrected by MLSC method (From Riess et al. 1998$^{\rm 17}$).}
 \label{riess}
  \end{center}
\end{figure*}

\section{SN Ia's role in cosmology}\label{sect:5}
The previous sections have discussed the properties of SNe Ia with
an emphasis on the corrections that can be applied to compensate
for the variations in the absolute luminosities and also in their
spectra.\footnote{When the apparent luminosities of objects at
differing redshifts are being compared, care must be taken to
correct for the redshifting of the spectral bands. This is known
as the so-called `K-correction'.} In this section we focus on how
equation (\ref{distance}) must be modified when the condition
$z\ll 1$ no longer holds. We emphasize how information concerning
the cosmological model can be extracted from the observed
redshift-apparent magnitude relation.

To this end we must first refine our definition of distance, which
is complicated by the expansion of the universe.  The `co-moving'
distance $d_{\rm CM}$ to an object at redshift $z$ is given by the
integral
 \begin{equation}
d_{\rm CM}=c{H_0}^{-1}\int _0^z{\rm d}z~ \left( \frac{ H_0 }{
H(z') } \right),
  \end{equation}
where $H$ is the Hubble constant, which is a function of $z$, and
$H_{\rm 0}$ is its present value. For a universe filled with
radiation, matter, and a cosmological constant
 \begin{equation}
\frac{ H_0 }{ H(z') } = \left[ \Omega _{\rm R}(1+z')^4+ \Omega
_{\rm M}(1+z')^3+ \Omega _{\rm k}(1+z')^2+ \Omega _{\rm \Lambda }
\right] ^{-1/2}
  \end{equation}
where $\Omega _{\rm k}=(1- \Omega _{\rm R} -\Omega _{\rm M}
-\Omega _{\rm \Lambda}).$ In the case of a `quintessence' or `dark
energy' component of constant $w=p/\rho c^{\rm 2}$ instead of a
cosmological constant (with $w=-1$), we would replace the $\Omega
_{\rm \Lambda}$ term with $\Omega _{\rm DE}(1+z')^{3(1+w)}.$ For
$\Omega _{\rm k}=0$ (a spatially flat universe), the `co-moving'
distance as defined above coincides with the `angular-diameter'
distance $d_{\rm AD},$ but in the presence of a nonflat spatial
geometry,

 \begin{equation}
d_{\rm AD}(z)=
\begin{cases}
{H_0}^{-1}(\Omega _{\rm k})^{-1/2}\sinh \left[ H_0(\Omega _{\rm k})^{1/2}d_{\rm CM}(z)\right] ,& \text{if $\Omega _{\rm k}>0,$}\\
d_{\rm CM}(z), & \text{if $\Omega _{\rm k}=0,$}\\
{H_0}^{-1}(-\Omega _{\rm k})^{-1/2}\sin \left[ H_0(-\Omega _{\rm
k})^{1/2}d_{\rm CM}(z)\right] , & \text{if $\Omega _{\rm k}<0.$}
\end{cases}
  \end{equation}
Finally, the `luminosity' distance $d_L$ is defined in terms of
the `angular-diameter' distance according to
 \begin{equation}
d_L(z)=(1+z)d_{\rm AD}(z)
  \end{equation}
where the $(1+z)$ factor has been inserted in order to account for
two effects: (1) as a photon propagates from redshift $z$ to us
today, its energy is redshifted, or diminished, by a factor of
$(1+z)^{-1},$ and (2) the rate of emission of photons is lowered
by a factor of $(1+z)^{-1}$ in terms of today's time, rather than
the time at the instant of emission.

It follows that using the formula
 \begin{equation}
\ell =\frac{L}{{d_L}^2}
  \end{equation}
where $L$ is the absolute (bolometric) luminosity and $\ell $ is
the apparent (bolometric) luminosity holds for all redshifts when
the luminosity distance $d_L$ as defined above is used.

The object of cosmological supernova studies is to measure the
function $d_L(z)$ beyond its first-order linear term whose
coefficient is $c{H_0}^{-1}.$ By measuring the higher-order
corrections (quadratic order and beyond), we uncover the equation
of state of what is driving the present expansion of the universe,
and through the cubic and higher terms discover how this equation
of state evolved with redshift. The SN Ia's role in cosmology has
been reviewed by many authors and further details on this
technique can be found in the these
reviews\cite{LEIBUNDGUT01,ASTIER12,KIM15,CARROLL92}. Here, we
summarize some historic events in this field.

In 1927, Lema\^{i}tre\cite{LEMAITRE1927} deduced that the velocity
of recession of distant objects relative to an observer can be
approximated to be proportional to distance, should Einstein's
general relativity equation be assumed to be true. In 1929,
Hubble\cite{HUBBLE1929} noticed that a linear relation actually
exists between an object's recession velocity and its distance,
judging by 22 external galaxies. This relation was henceforth
named Hubble's law, and hallmarks the discovery of the expansion
of the universe. After this discovery, two questions arise
naturally, i.e. what is the future of the universe, and what are
the energy sources driving the expansion of the universe. To
answer these questions, one needs to measure the recession
velocity variation as a function of time or redshift, i.e. the
deceleration parameter of the expansion $q_{\rm 0}$. In modern
cosmology, including Einstein's cosmological
constant\cite{EINSTEIN1917}, the cosmic deceleration parameter is
defined as

 \begin{equation}
q_{\rm 0}=\frac{\Omega_{\rm M}}{2}-\Omega_{\rm \Lambda},
\label{q0}
  \end{equation}
where $\Omega_{\rm M}$ is the ratio between the actual mass
density of the universe and the critical density above which the
universe would collapse again someday if $\Omega_{\rm \Lambda}=0$,
which is only dependent on the actual mass density of the
universe, and $\Omega_{\rm \Lambda}$ describes how the
cosmological constant affects the expansion of the
universe\cite{CARROLL92}. For a universe of $\Omega_{\rm
\Lambda}\neq0$, the behavior of the universe is determined by
$\Omega_{\rm M}$ and $\Omega_{\rm \Lambda}$ together. $q_{\rm
0}=0$ leads to a constant expansion velocity, $q_{\rm 0}<0$ an
accelerating expansion, and $q_{\rm 0}>0$ a decelerating
expansion. It can be clearly seen in the equation that the
expansion of the universe is accelerating if $\Omega_{\rm
\Lambda}>\Omega_{\rm M}/2$. For a universe of $\Omega_{\rm
\Lambda}=0$, $q=\Omega_{\rm M}/2$, and therefore, the universe is
always decelerating, where $q>\frac{1}{2}$, $q=\frac{1}{2}$ and
$q<\frac{1}{2}$ corresponds to the closed, flat and open universe,
respectively (please see Fig. \ref{riess}).

Although it was known that SNe Ia are very good standard
candles\cite{BRANCH92}, significant progress on this subject had
yet to be made until the 1990s, after the advent of CCD
imagers\cite{HAMUY96b,HAMUY96c}. In the 1990s, two teams focused
on the measurement of the distance to high-z supernovae, i.e. the
Supernova Cosmology Project (SCP) and the High-z team (HZT). Both
teams built up large samples, including those measured by the
Hubble Space Telescope (HST), and then individually published
their twin studies\cite{RIESS98,PERLMUTTER99}. In the twin
studies, they adjusted the parameters of an ($\Omega_{\rm M}$,
$\Omega_{\rm \Lambda}$) universe to fit their data, as shown in
Fig. \ref{riess} which is an example from Riess et
al.\cite{RIESS98}, and obtained a similar conclusion. The
surprising result is that high-z SNe Ia were observed to be
fainter than expected from their low-z counterparts in a
matter-dominated universe. A matter-dominated closed universe
(i.e. $\Omega_{\rm M}=1$) was unambiguously ruled out at a
confidence level of greater than 7 $\sigma$. Neither does an open,
$\Lambda=0$ cosmology fit the data well; the cosmological constant
is nonzero and positive, with a confidence level of
P($\Lambda>0$)=99\%. For a flat universe ($\Omega_{\rm M}$ +
$\Omega_{\rm \Lambda}$=1), a positive $\Omega_{\rm \Lambda}$ was
found to be required at a confidence level of 7 $\sigma$.
Actually, the probability of an accelerating expansion of the
universe with ($\Omega_{\rm M}>0$, $\Omega_{\rm \Lambda}>0$,
$q_{\rm 0}<0$) was modest, typically 3 $\sigma$ or less, i.e. the
confidence level for a $\Lambda$ dominated accelerating expansion
was less than 3 $\sigma$. For a flat universe, the best-fit
parameters are $\Omega_{\rm M}\approx0.3$ and $\Omega_{\rm
\Lambda}\approx0.7$, which supported an accelerated expansion as
derived from equation (\ref{q0}). The twin studies are regarded as
the original evidence for the discovery of the
accelerating-expansion universe, and in 2011, Saul Perlmutter,
Brian Schmidt and Adam Riess were awarded the Nobel Prize in
Physics for the discovery. It should be noticed that the expansion
of the universe is actually evolving. At $z\approx0.45$, the
density of matter and dark energy was equal, and matter dominated
at early epochs, i.e. the expansion of the universe was
decelerating at $z>0.45$\cite{ASTIER12}.

The discovery of accelerating expansion today implies that our
universe is dominated by some unknown form of energy, which is
often characterized by the ratio of pressure to density, i.e. the
dark energy equation of state, $w=P/\rho c^{\rm 2}$. For some
arbitrary length scale $a$, the density of the universe evolves as
 \begin{equation}
\rho\propto a^{-3(1+w)}. \label{a}
  \end{equation}
From this equation, it is easily seen that $w=0$ for normal
matter, i.e. the density decreases as the universe expands. If
$w=-1$, which corresponds to Einstein's cosmological constant, its
density is a constant, and does not dilute with the expansion of
the universe. A value of $w<-1$ is permitted although it is
debatable theoretically. In this situation, the dark energy
density will increase with time, and may ultimately lead to the
destruction of galaxies\cite{HOWELL11,FRIEMAN08}. If $-1<w<0$, the
dark energy density will decrease with time, and it will become
problematic whether the expansion of the universe will continue to
accelerate if the value of $w$ is large enough. Were $\Omega_{\rm
M}$ assumed to be negligible, a value of $w<-\frac{1}{3}$ for dark
energy would be sufficient for the indefinite acceleration of its
expansion. However, considering that it is very likely for
$\Omega_{\rm M}$ to have a value of larger than
0.2\cite{RIESS98,PERLMUTTER99}, the dark energy has to have an
even smaller value of $w$ for this to happen. It is verified that
the critical point for the expansion of the universe to continue
accelerating is in fact $w=-\frac{1}{2}$\cite{KNOP03}. However, it
is possible that $w$ is not a constant, but rather a variable
which also evolves with time in the general form $w(z)=w_{\rm
0}+w_{\rm a}z/(1+z)$, where $w$ will be a constant $w_{\rm 0}$ if
$w_{\rm a}=0$.

\begin{figure*}
\begin{center}
\includegraphics[totalheight=3.in,width=4.in,angle=0]{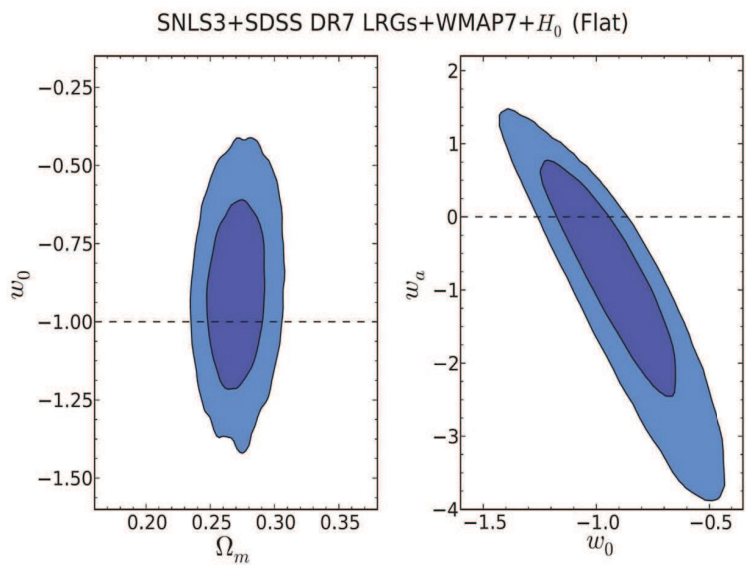}
 \caption{Confidence level contours in the plane of cosmological parameters
 ($\Omega_{\rm m}$, $w_{\rm 0}$, $w_{\rm a}$). The horizontal dashed lines indicate a cosmological constant,
 i.e. $w_{\rm 0}=-1$ and $w_{\rm a}=0$ (From Fig. 6 in Sullivan et al.
2011$^{\rm 120}$).}
 \label{sullivan2}
  \end{center}
\end{figure*}

Following the discovery of dark energy, the equation of the state
of dark energy (EOSDE) and the evolution of EOSDE with time became
burning issues. Although the measurement of $w$ is still
consistent with an accelerating expansion of the universe, the
uncertainty of $w$ is still very large in those early
observations. The value of $w$ was generally located between
$-0.7$ and $-1.6$\ at a typical confidence level of
95\%\cite{TONRY03,KNOP03,RIESS04}. Today, more and more evidence
based on SNe Ia observations from SNe Ia groups and other
observational constraints, such as those from Supernova Legacy
Survey (SNLS), Harvard-Smithsonian Center for Astrophysics (CfA),
Sloan Digital Sky Survey (SDSS) and SCP, supports a flat universe
with a constant $w=-1$, i.e. at least rapidly evolving forms of
dark energy were ruled out although the current observational
constraints on $w_{\rm a}$ are
weak\cite{KESSLER09,HICKEN09,AMANULLA10,RIESS07,MIKNAITI07,WOODVASEY07,CONLEY11,SULLIVAN11}.
For example, in Fig. \ref{sullivan2}, combined with full Wilkinson
Microwave Anisotropy Probe 7-year (WMAP7) power spectrum, SNLS3
and SDSS DR7 data, $\Omega_{\rm M}=0.271\pm0.015$, $w_{\rm
0}=-0.905\pm0.196$ and $w_{\rm a}=-0.984^{+1.094}_{-1.097}$ were
obtained, which are consistent with a flat, $w=-1$
universe\cite{SULLIVAN11}. For most of these studies, the HST
played a vital role in determining the cosmological parameters,
and provides many high-quality, high-z observation which remain
one of our most efficient tools for understanding the properties
of dark energy\cite{RIESS04,RIESS07,GARNAVICH98}. An event at
$z=1.914$ was even discovered recently by HST, whose magnitude is
consistent with that expected from $\Lambda$CDM
cosmology\cite{JONES13}.

\section{Issues and prospects}\label{sect:6}
SNe Ia are now applied to constrain the properties of the dark
energy and the evolution of the properties with time. Although SNe
Ia may provide more strict constraints than some other surveys,
such as baryon acoustic oscillations (BAO) and weak gravitational
lensing (WL), the combination of the results from more than one
survey is often applied to constrain the EOSDE
parameter\cite{FRIEMAN08,SULLIVAN11}. For dark energy, a change of
1\% in $w$ only leads to a change of 3\% in dark energy density at
redshift $z = 2$ and a change of 0.2\% at $z=1$ -
$2$\cite{FRIEMAN08}. Therefore, to study the properties of the
dark energy quantitatively, a precision of better than 2\% is
required\cite{HOFLICH10}. However, even for the most well-observed
sample, the present distance accuracy seems not to be good enough.
At present, a well-observed low redshift sample may fulfill the
requirement of 2\% precision, but the distance error increases to
$\sim$ 5\% at redshifts greater than 1\cite{KIM15,ASTIER14}. As we
know, SNe Ia with redshifts greater than 1 provide the best
constraints on the EOSDE.

For precision cosmology, there are still several difficulties with
regard to SNe Ia. Firstly, not all SNe Ia may be taken to be
distance indicators, such as 2002cx-like supernovae\cite{LIWD03}.
Such SNe Ia conform to no relations which regular SNe Ia do, such
as the Phillips relation, and therefore cannot be standardised.
Another sub-class of SNe Ia that do not adhere to the Phillips
relation are the so-called super-luminous SNe Ia\cite{HOWELL06}.
If these peculiar supernovae are not excluded from a SNe Ia
sample, the confidence level of the resulting distance measurement
would decrease greatly. Secondly, new light curve analysis methods
which have been developed tend to be based on new intrinsic
parameters, which are assumed either to be invariant with
redshift, or at least to vary with redshift in such a way that
does not introduce further systematic uncertainty. Despite the
reduction of the luminosity uncertainty from 0.15 mag to less than
0.10 mag, or better than 5\% in terms of distance
uncertainty\cite{MANDEL11,KIM13}, which these methods serve to
achieve, such assumptions ought to be checked carefully, since it
has been shown that systematic-uncertainty-inducing evolution may
exist\cite{HOWELL07,SULLIVAN09}, and besides the light-curve shape
and the color, a third parameter correlated to one or some
properties of the host galaxy is necessary. Finally, it is crucial
to increase the detection efficiency of SNe Ia at high redshifts
for further studies, especially at redshifts greater than 1. Two
factors may affect this detection efficiency. One is that only
luminous SNe Ia are likely to be detected due to Malmquist bias,
but luminous SNe Ia are relatively rare. The other is that most of
the radiation may be shifted into infrared bands for the SNe Ia at
high redshifts, which makes it difficult to detect SNe Ia by
optical telescopes. Infrared space telescopes, in this context,
may be an alternative for future SNe Ia
science\cite{LEIBUNDGUT01}. For example, the recently installed
Wide-Field Camera 3 on HST has granted SN surveys an unprecedented
depth, i.e. it is possible to detect a typical SN Ia at
$z\approx2.5$ by HST. Another merit of infrared observations is
that the intrinsic dispersion of SNe Ia after correction is
smaller than optical band as shown in Fig. \ref{phillips1}. In
addition to the difficulties mentioned above, for high-z
supernovae, the confidence level for classification by spectrum is
not as high as that for low-z supernovae. For a high-z SNe Ia
sample, although practical studies of photometric identification
have developed methods which achieve a purity of $\sim95$\% for
SNe Ia\cite{KESSLER10,SAKO11,BAZIN11}, there exists a possibility
that such samples may be contaminated by a few type Ib/c
supernovae\cite{JONES13,RODNEY12}. Therefore, techniques to remove
such sample impurities need to be developed.

However, if an appropriate observational strategy is adopted, the
effect of time dilation could provide a great opportunity for
high-z SN observation. If the cosmological redshift is derived
from the expansion of the universe, the observed time interval of
an event will be dilated by a factor of $(1+z)$, which means that
the light curve of a high-z SN Ia will present a broadening peak
and a more shallow slope for the later exponential decline phase.
Therefore, the effect of time dilation increases the probability
to detect the high-z SNe Ia, although it becomes relatively
difficult to obtain a complete light curve.

\section*{Acknowledgments}
We are very grateful to the anonymous referees for their kind
comments that improved this manuscript greatly. This work was
partly supported by NSFC (11473063,11522327, 11390374, 11521303),
the Western Light Key Project of the Chinese Academy of Sciences
and Key Laboratory for the Structure and Evolution of Celestial
Objects, Chinese Academy of Sciences. Z.H. thanks the support by
the Strategic Priority Research Program ``The Emergence of
Cosmological Structures" of the Chinese Academy of Sciences, Grant
No. XDB09010202, and Science and Technology Innovation Talent
Programme of the Yunnan Province (Grant No. 2013HA005).



\end{document}